\documentclass[english]{paper}
\usepackage{mathptmx}

\usepackage[T1]{fontenc}
\usepackage[latin9]{inputenc}
\usepackage{amssymb}
\usepackage{esint}
\usepackage[numbers]{natbib}

\makeatletter
\newcommand{\lyxaddress}[1]{
\par {\raggedright #1
\vspace{1.4em}
\noindent\par}
}

\def\eg{{ e.g.,\ }}
\def\ie{{ i.e.,\ }}
\usepackage{hyperref}

\makeatother

\usepackage{babel}

\begin{document}

\title{On the Structure and Scale of Cosmic Ray Modified Shocks%
\thanks{accepted for publication in PPCF%
} }

\author{M A Malkov$^{1}$, P H Diamond$^{1}$ and R Z Sagdeev$^{2}$}

\maketitle

\lyxaddress{$^{1}$CASS and Department of Physics, University of California,
San Diego, La Jolla, CA 92093}

\lyxaddress{$^{2}$University of Maryland, College Park, Maryland 20742-3280,
USA}

mmalkov@ucsd.edu
\begin{abstract}
Strong astrophysical shocks, diffusively accelerating cosmic rays
(CR) ought to develop CR precursors. The length of such precursor
$L_{p}$ is believed to be set by the ratio of the CR mean free path
$\lambda$ to the shock speed, i.e., $L_{p}\sim c\lambda/V_{sh}\sim cr_{g}/V_{sh}$,
which is formally independent of the CR pressure $P_{c}$. However,
the X-ray observations of supernova remnant shocks suggest that the
precursor scale may be significantly shorter than $L_{p}$ which would
question the above estimate unless the magnetic field is strongly
amplified and the gyroradius $r_{g}$ is strongly reduced over a short
(unresolved) spatial scale. We argue that while the CR pressure builds
up ahead of the shock, the acceleration enters into a strongly nonlinear
phase in which an acoustic instability, driven by the CR pressure
gradient, dominates other instabilities (at least in the case of low
$\beta$ plasma). In this regime the precursor steepens into a strongly
nonlinear front whose size scales with \emph{the CR pressure }as $L_{f}\sim L_{p}\cdot\left(L_{s}/L_{p}\right)^{2}\left(P_{c}/P_{g}\right)^{2}$,
where $L_{s}$ is the scale of the developed acoustic turbulence,
and $P_{c}/P_{g}$ is the ratio of CR to gas pressure. Since $L_{s}\ll L_{p}$,
the precursor scale reduction may be strong in the case of even a
moderate gas heating by the CRs through the acoustic and (possibly
also) the other instabilities driven by the CRs.\end{abstract}

\section{Introduction}

Significant progress in observations of galactic supernova remnant
shocks (SNR) over the passed decade have furthered our understanding
of the particle acceleration mechanism which is deemed responsible
for both the individual SNR emission and for the galactic cosmic ray
(CR) production as a whole. The mechanism is known as the first order
Fermi or diffusive shock acceleration (DSA, see \citep{BlandEich87}
for a comprehensive and \citep{KirkDendy01} for short review). The
shocks are typically observed as surprisingly thin filaments, particularly
well resolved in the X-ray band. The working hypothesis is that this
emission is due to the super-TeV shock accelerated, synchrotron radiating
electrons, occasionally also seen in gamma-rays (through the inverse
Compton (IC) up-scattering of the background photons). The gamma emission
may in some cases be contaminated or even dominated by the accelerated
protons via $\pi^{0}$ decay \citep{Abdo10W44short,AbdoIC443_10short}.
A convincing demonstration of namely the latter scenario would, of
course, be a \emph{prima facie }evidence for the acceleration of also
the main, i.e. hadronic component of galactic CRs in SNRs. 

The key element of the DSA is multiple crossing of the shock front
with a $\sim V_{sh}/c$ energy gain after each crossing. In doing
so particles diffusively escape from the shock, on average to a distance
$L_{p}\sim\kappa\left(p\right)/V_{sh}$. Here $\kappa$ is the momentum
dependent diffusion coefficient and $V_{sh}$ is the shock velocity.
An obvious morphological consequence of this process should be and
extended $\sim L_{p}$ shock precursor filled with synchrotron radiating
electrons. A disturbing fact of the high-resolution $X$-ray observations
is that there is no supporting evidence for such extended precursors
\citep{LongRaymond03,Bamba05,Bamba03,BalletRev06}. Note that $L_{p}$
grows with particle momentum as almost certainly does $\kappa\left(p\right)$,
so particles of higher energy should make thicker emission filaments
than do the lower energy particles. This trend does not seem to be
supported by the observations either, see \citep{Bamba03,CassamChenHughes07}.

According to another widely accepted view, the particle diffusion
coefficient $\kappa$ should be close to the Bohm value, $\kappa\sim cr_{g}\left(p\right)/3$,
which requires strong magnetic fluctuations $\delta B_{k}\sim B_{0}$
at the resonant scale $k\sim1/r_{g}\left(p\right)$. The high level
of fluctuations is achieved through one of the instabilities driven
by accelerated particles. The following three CR driven instabilities
have been suggested to generate magnetic field fluctuations. The first
one is the well known \citep{SagdShafr61} ion cyclotron resonant
instability of a slightly anisotropic (in pitch angle) CR distribution
ahead of the shock \citep{Bell78}. The free energy source of this
instability has the potential to generate very strong field fluctuations
\citep{McKVlk82} 

\begin{equation}
\left(\delta B/B_{0}\right)^{2}\sim M_{A}P_{c}/\rho V_{sh}^{2}.\label{delB}\end{equation}
where $M_{A}\gg1$ is the Alfvenic Mach number, $P_{c}$ is the CR
pressure, $\rho$ is the gas density and $u_{s}$ is the shock velocity.
However, the actual turbulence level was thought to remain moderate,
$\delta B\sim B_{0}$(\eg \citep{McKVlk82,AchtBland86}).

The second instability, is a nonresonant instability driven by the
CR current. The advantage of this instability seems to be twofold.
First, it cannot be stabilized by the quasilinear deformation of the
CR distribution function since in the upstream plasma frame the driving
CR current persists, once the CR cloud is at rest in the shock frame.
Second, it generates a broad spectrum of waves, and the longest ones
were claimed to be stabilized only at the level $\delta B\gg B_{0}$,
due to the lack of an efficient stabilization mechanism at such scales.
Within the context of the CR acceleration, this instability was first
studied by \citet{Acht83FireHose}, but the potential of this instability
to generate strong magnetic fields was first emphasized in refs. \citet{BelLuc01,Bell04},
so the instability is often referred to as the Bell's instability.
The stabilization mechanism was assumed to be the magnetic tension
\citep{Bell04}. Bell also pointed out that in the most interesting
regime, the instability is driven by a fixed CR return current through
the Ampere force $\mathbf{J}_{c}\times\mathbf{B}$. It should be noted,
however, that the dissipation of the return current due to the anomalous
resistivity still needs to be addressed.

The third instability is an acoustic instability (also known as Drury's)
driven by the pressure gradient of accelerated CRs upstream \citep{DruryFal86}.
This instability was also studied numerically by Dorfi \citep{Dorfi84}.
The pressure gradient is clearly a viable source of free energy for
the instability. So, among quantities varying across a shock, the
pressure jump is the most pronounced one in that sense that it does
not saturate with the Mach number, unlike the density or velocity
jumps. 

Curiously enough, the acoustic instability has received much less
attention than the first two. Moreover, in many numerical studies
of the CR shock acceleration, special care is taken to suppress it.
The suppression is achieved by using the fact that a change of stability
occurs at that point in the flow where $\partial\ln\kappa/\partial\ln\rho\simeq-1$
(for both stable and unstable wave propagation directions). Here $\kappa$
is the CR diffusion coefficient, and $\rho$ is the gas density. Namely,
one requires this condition to hold identically all across the shock
precursor, \ie where the CR pressure gradient $\nabla P_{c}\neq0$.
Not only is this requirement difficult to justify physically, but,
more importantly, an \emph{artificial} suppression of the instability
eliminates its \emph{genuine }macroscopic and microscopic consequences,
as briefly discussed below. 

Among the macroscopic consequences  an important one is the vorticity
generation through the baroclinic effect (missalignment of the density
and pressure gradients $\nabla\rho\times\nabla P\neq0$, e.g. \citep{KulsrudMF97}).
Here $\nabla P$ may be associated with a quasi-constant macroscopic
CR-gas pressure gradient $\nabla P_{c}$, generally directed along
the shock normal. Variations of $\nabla\rho$ are locally decoupled
from $P_{c}$, unlike in the situation in a gas with a conventional
equation of state where $P=P\left(\rho\right)$ and where the baroclinic
term vanishes. The vorticity generation obviously results (just through
the frozen in condition) in magnetic field generation, so that the
field can be amplified by the CR pressure gradient. More importantly,
this process amplifies the \emph{large scale field}, required for
acceleration of \emph{high energy particles}. Furthermore, the amplification
takes place well ahead of the gaseous subshock. The both requirements
are crucial for improving high energy particle confinement and making
the shock precursor shorter in agreement with the observations. Large
scales should be present in the ambient plasma as a seed for their
amplification by the acoustic instability and could be driven/seeded
by wave packet modulations. Apart from that, they result from the
coalescence of shocks formed in the instability, and from the scattering
of Alfven waves in $k$-space by these shocks to larger scales \citep{DM07}.
Note that the Bell instability is essentially a short scale instability
(the maximum growth rate is at scales much smaller than the gyro-radii
of accelerated particles). At larger scales the growth rate decreases
and the instability transitions into the conventional cyclotron instability.
It should be noted that, as well known \citep{Kevlahan97,KulsrudMF97,GiacJok07},
vorticity (and thus magnetic field) can be efficiently generated also
at the subshock. This would, of course, be too late for improving
particle confinement and reducing the scale of the shock precursor.

Now the question is which instability dominates the CR dynamics? Given
the finite precursor crossing time, it is reasonable to choose the
fastest growing mode and consider the development of a slower one
on the background created by the fast mode after its saturation. The
resonant cyclotron instability is likely to dominate at the outskirt
of the shock precursor where both the CR current and pressure gradient
(driving the other two instabilities) are weak, whereas the pitch
angle anisotropy is strong enough to drive the resonant instability.
Recall that the anisotropy is typically inversely proportional to
the local turbulence level which must decrease with the distance from
the shock. In this paper, however, we focus on the main part of the
shock precursor where both the CR-pressure gradient and CR current
are strong, and consider the Bell and Drury instabilities as the strongest
candidates to govern the shock structure. In fact, these instabilities
are coupled, not only by the common energy source but also dynamically.
The coupling of the magnetic and the density perturbations, akin to
the modulational instability will be the subject of a separate publication.
This paper will be limited to a simpler situation in which one of
the instabilities dominates. In particular we will identify conditions
under which the acoustic instability grows more rapidly. Then, we
determine the shock structure resulting from the nonlinear development
of the acoustic instability.

\section{Comparison of the growth rates}

The derivations of the linear dispersion relations of the Bell and
Drury modes can be found in refs.\citep{DruryFal86,Bell04}. For the
magnetic field and density perturbations of the form, $\tilde{B}\propto\exp\left(-ikx\pm i\omega t\right)$
and $\tilde{\rho}\propto\exp\left(ikx\pm i\omega t\right)$, in an
approximate symmetric representation, in which the diffusive CR damping
of acoustic mode and the resonant CR contribution to the nonresonant
magnetic instability are neglected, these relations can be written
as follows

\begin{eqnarray}
\omega^{2} & =k^{2}C_{A}^{2} & -2\gamma_{B}kC_{A}\label{eq:Bdisp}\\
\omega^{2} & =k^{2}C_{s}^{2} & -2i\gamma_{D}kC_{S}\label{eq:Ddisp}\end{eqnarray}
Here $C_{A}$ and $C_{S}$ are the Alfven and the sound speeds, respectively
while

\begin{equation}
\gamma_{B}=\sqrt{\frac{\pi}{\rho_{0}}}J_{c}/c\;\;\mathrm{and}\;\gamma_{D}=-\frac{1}{2\rho_{0}C_{S}}\frac{\partial\overline{P}_{c}}{\partial x}\label{eq:BDgrowthrates}\end{equation}
The last expressions approximately represent the maximum growth rates
of the nonresonant B-field (Bell) perturbations and acoustic density
(Drury) perturbations, driven by the CR return current and by the
CR pressure gradient, respectively. Note that the instabilities are
different in that the Bell mode is unstable in the limited $k$-band
where $0<k<2\gamma_{B}/C_{A}$, and $\gamma_{B}$ is only the maximum
growth rate that is obviously achieved at $k=\gamma_{B}/C_{A}$, whereas
the acoustic mode saturates with $k$ at $k\gtrsim\gamma_{D}/C_{S}$
at the level $\gamma_{D}$. 

Next, we compare the growth rates of the two instabilities. The comparison
should be done for the same equilibrium distribution of CRs upstream
of the subshock, \ie in the CR precursor. Apparently, there is a
difficulty here. The acoustic instability analysis presumes a certain
level of magnetic fluctuations (e.g. $\delta B\sim B$) to pondermotively
\citep{AchterPonder81} couple CRs to the gas flow, and thus ensure
the equilibrium. In the Bell's stability analysis, it is assumed that
the undisturbed B-field is along the shock normal and $\mathbf{J_{\mathbf{CR}}}$
points into the same direction. Then, the CR pressure gradient cannot
be statically compensated since $\mathbf{J_{\mathbf{CR}}}\times\mathbf{B}=0$.
Implicitly though, one may assume that the resonant instability provides
Alfven waves which balance the CR pressure gradient along the main
field through the wave pondermotive pressure at larger scales. Then,
the Bell instability develops at the scales much shorter than the
gyro-radii of the current-carrying particles. Note that such treatment
cannot be extended to the larger scales without further considerations
\citep{DM07,BykovBell09}. In a strongly modified shock, the equilibrium
distribution upstream in a steady state is given by \citep{m97a}:

\begin{equation}
f=f_{0}\left(p\right)\exp\left[\frac{q\left(p\right)}{3\kappa\left(p\right)}\phi\left(x\right)\right],\;\; x\ge0\label{eq:fSteadySol}\end{equation}
where $\phi$ is the flow potential, $u=\partial\phi/\partial x<0$,
$f_{0}\left(p\right)$ is the CR distribution and $q\left(p\right)=-\partial\ln f_{0}/\partial\ln p$
the spectral index at the subshock. On the other hand, if the shock
modification is negligible, the equilibrium is simply

\begin{equation}
f=f_{0}\left(p\right)\exp\left[\frac{1}{\kappa\left(p\right)}\phi\left(x\right)\right],\;\; x\ge0\label{eq:fSteadySolTP}\end{equation}
One sees that the only difference between the last two representations
of the particle distribution upstream is the $q/3$ factor in the
exponent of eq.(\ref{eq:fSteadySol}). This quantity is well constrained
in the nonlinear solution given by eq.(\ref{eq:fSteadySol}): $3.5<q\left(p\right)<q_{sub}$,
where $q_{sub}=3r_{sub}/\left(r_{sub}-1\right)$ with $r_{sub}$ being
the subshock compression ratio. For the ratio of acoustic to magnetic
growth rates, eq.(\ref{eq:BDgrowthrates}), we obtain

\begin{equation}
\frac{\gamma_{D}}{\gamma_{B}}=\frac{C_{A}}{C_{S}}\frac{c^{2}}{3\omega_{ci}}\left\langle \frac{p^{2}}{\sqrt{1+p^{2}}}\frac{q}{3\kappa}\right\rangle \label{eq:GrRateRat1}\end{equation}
where $p$ is in units of $mc$. We have introduced the spectrum averaged
quantity above as

\begin{equation}
\left\langle \cdot\right\rangle \equiv\frac{\int\left(\cdot\right)f_{0}\left(p\right)p^{2}\exp\left(q\phi/3\kappa\right)dp}{\int f_{0}\left(p\right)p^{2}\exp\left(q\phi/3\kappa\right)dp}\label{eq:AverDef}\end{equation}
and where for the test particle solution, given by eq.(\ref{eq:fSteadySolTP})
one should replace $q/3\to1$. For the Bohm diffusion coefficient
$\kappa=r_{g}\left(p\right)c/3$, we obtain

\begin{equation}
\frac{\gamma_{D}}{\gamma_{B}}=\left\langle \frac{q}{3}\right\rangle \frac{C_{A}}{C_{S}}\label{eq:GrRateRatio2}\end{equation}
From the last formula we may conclude that the acoustic instability
dominates the magnetic one in the case of low $\beta\equiv C_{S}^{2}/C_{A}^{2}\ll1$
upstream, regardless of the degree of nonlinearity of acceleration.
The reason for such a counter-intuitive relation between the growth
rates of these two instabilities is that a stronger magnetic field
(low-beta plasma) supports a stronger CR pressure gradient which drives
the acoustic instability. Note, that plasma heating upstream would
increase the role of magnetic instability unless the large scale magnetic
field is also strengthened, e.g. through an inverse cascade of the
turbulent magnetic energy \citep{DM07}. On the other hand, the development
of the acoustic instability makes (Sec.\ref{sec:Turbulent-CR-Front})
the precursor shorter. This boosts the gradient driven acoustic instability
and leaves less room for the current driven magnetic instability.
Next, we consider the CR transport in a developed acoustic turbulence
with an admixture of Alfven waves, presumably generated by cyclotron
instability at a distant part of the shock precursor and convected
into its core.

\section{CR transport in Shock Precursor\label{sec:CR-transport-in}}

We assume that magnetic perturbations in the precursor are of the
following two types. First, there are conventional shear Alfven waves,
stemming from the CR cyclotron instability. Second, there are compressible
magnetosonic perturbations generated by the Drury instability. The
pitch angle scattering of CRs in the both wave fields has been calculated
in many publications. We can use the expressions given by eqs.(6,7)
in ref.\citep{ChandranGS00PhRvL}. After summing the Bessel function
series and retaining only the magnetic parts of the scattering wave
fields, we obtain for the pitch-angle diffusion coefficients the following
expressions

\begin{eqnarray}
D_{\mu}^{A} & = & -\left(1-\mu^{2}\right)\sum_{\mathbf{k}}\frac{1}{\xi^{2}}\intop_{0}^{\infty}I^{A}\left(k_{\parallel},k_{\perp},\tau\right)\nonumber \\
 & \times & \cos\left(k_{\parallel}c\mu\tau\right)d\tau\frac{\partial^{2}}{\partial\tau^{2}}J_{0}\left(2\xi\sin\frac{\Omega\tau}{2}\right)\label{eq:DA}\end{eqnarray}

\begin{eqnarray}
D_{\mu}^{S} & = & \frac{1}{3}\left(1-\mu^{2}\right)\sum_{\mathbf{k}}\frac{1}{\xi^{2}}\intop_{0}^{\infty}I^{S}\left(k_{\parallel},k_{\perp},\tau\right)\frac{k_{\parallel}^{2}}{k_{\perp}^{2}}d\tau\nonumber \\
 & \times & \left[\Omega^{2}\cos\left(k_{\parallel}c\mu\tau\right)-\xi^{-2}\frac{\partial^{2}}{\partial\tau^{2}}\right]J_{0}\left(2\xi\sin\frac{\Omega\tau}{2}\right).\label{eq:DS}\end{eqnarray}
Here $I^{A}$ and $I^{S}$ are the normalized (to $B_{0}^{2}/8\pi$)
spectral densities of magnetic fluctuations of Alfven and slow magnetosonic
wave components of the upstream turbulence, $\mu$ is the cosine of
particle pitch angle, $\xi=k_{\perp}c\sqrt{1-\mu^{2}}/\Omega$, $J_{0}$
is the Bessel function, and $\Omega$ is the relativistic gyrofrequency.
The spatial diffusion coefficient can be evaluated as follows (e.g.\citep{Jokipii66})

\begin{equation}
\kappa=\frac{c^{2}}{8}\left\langle \frac{1-\mu^{2}}{\mathcal{D}_{\mu}}\right\rangle \label{eq:kappa}\end{equation}
where $\mathcal{D}_{\mu}=\left(D_{\mu}^{A}+D_{\mu}^{S}\right)/\left(1-\mu^{2}\right)$
and $\left\langle \cdot\right\rangle $ denotes here the pitch angle
averaging. For the turbulence spectra extended in $k_{\perp}$ (such
as the Goldreich-Shridhar cascade \citep{Goldr95}), the scattering
frequency $\mathcal{D_{\mu}}$ peaks at $\left|\mu\right|=0,1$ \citep{ChandranGS00PhRvL}.
Indeed, the scattering is known to be strongly suppressed for $\xi\gg1$
(high frequency  perturbation of particle orbits), so that particles
with $\left|\mu\right|\approx1$ are subjected to a more coherent
(not so rapidly oscillating) wave field. Furthermore, particles with
$\mu\approx0$ are effectively mirrored by the compressible component
of magnetic turbulence. Once there are peaks at $\left|\mu\right|=0,1$,
$\mathcal{D_{\mu}}$ should also have minima in between, i.e. at some
$\left|\mu\right|=\mu_{0}$, where $0<\mu_{0}<1$. Upon writing

\[
\left\langle \frac{1-\mu^{2}}{\mathcal{D}_{\mu}}\right\rangle =\intop_{0}^{1}\left(1-\mu^{2}\right)d\mu\intop_{0}^{\infty}\exp\left(-\mathcal{D}_{\mu}t\right)dt\]
and evaluating the integral in $\mu$ by using the steepest descent
method, we obtain

\[
\kappa=\frac{2\pi c^{2}\left(1-\mu_{0}^{2}\right)}{\sqrt{\mathcal{D}_{\mu}\left(\mu_{0}\right)\mathcal{D}_{\mu}^{\prime\prime}\left(\mu_{0}\right)}}\]
where the double prime denotes the second derivative. Assuming that
the compressible part of the turbulence, which originates from the
acoustic instability, dominates ($D^{A}\ll D^{S}$), we introduce
the the following momentum averaged diffusion coefficient, that will
be used in the next section

\begin{equation}
\bar{\kappa}=\frac{\int\kappa\left(p\right)f\left(p\right)p^{3}dp}{\int f\left(p\right)p^{3}dp}=\frac{\bar{\kappa}_{B}}{F_{S}+\alpha}\label{eq:kappabar}\end{equation}
Here $\bar{\kappa}_{B}$ is the momentum averaged Bohm diffusion coefficient
$\kappa_{B}=cr_{g}\left(p\right)/3$, $\alpha$ is the normalized
level of Alfvenic turbulence $\alpha\sim\left(\delta B_{A}\right)^{2}/B_{0}^{2}$,
(originating from $D^{A}$), and $F_{S}$ is the level of compressible
turbulence (originating from $D^{S}$). The latter can be calculated
in the simplest 1D model in which the acoustic waves unstably grow
at a rate $\gamma_{D}$ and then steepen into an ensemble of shocks
(shocktrain) \citep{MD06,MD09}

\begin{equation}
F_{S}=\sigma\left(\frac{\partial P_{c}}{\partial x}\right)^{2}\;\;\;\;\mathrm{with}\;\;\quad\sigma=\left(\frac{L_{s}}{\rho C_{S}^{2}}\right)^{2}\mathcal{F}\left(\vartheta_{nB}\right)\label{eq:FsSigma}\end{equation}
Here $L_{s}$ is the average distance between the shocks in the ensemble,
and $\mathcal{F}\sim1$ depends predominantly on the shock obliquity
and on the specific model for the shocktrain formation \citep{MalkKennel91,MedvDiam96}.

\section{Turbulent CR Front Solution\label{sec:Turbulent-CR-Front}}

In this section we derive the equations for the shock transition from
the kinetic convection-diffusion equation that can be written in a
steady state as follows

\begin{equation}
\frac{\partial}{\partial x}\left(ug+\kappa\frac{\partial g}{\partial x}\right)=\frac{1}{3}\frac{\partial u}{\partial x}p\frac{\partial g}{\partial p}\label{eq:ConvDif}\end{equation}
where $g=fp^{3}$ and $u$ is the (positive) magnitude of the flow
speed. We assume that the CR pressure integral is dominated by particles
with $p\gg mc$, but there is a spectral break at some $p=p_{*}$
which makes the integral finite without introducing a cut-off momentum
\citep{MD06}. This makes the well known two-fluid model \citep{Drury81}
suitable for our purposes. More problematic is the introduction of
$\bar{\kappa}_{B}$, eq.(\ref{eq:kappabar}), which may require the
upper cutoff. Integrating the appropriately weighted convection-diffusion
equation (\ref{eq:ConvDif}) in momentum, and supplementing the result
with the conservation of the mass and momentum fluxes across the shock
transition, we arrive at the following closed system of equations
that governs the shock transition

\[
\frac{\partial}{\partial x}\left(uP_{c}+\frac{\bar{\kappa}_{B}}{\sigma\left(\partial P_{c}/\partial x\right)^{2}+\alpha}\frac{\partial P_{c}}{\partial x}\right)=-\frac{1}{3}\frac{\partial u}{\partial x}P_{c}\]

\[
\rho u=\rho_{1}u_{1}=const\]

\begin{equation}
\rho u^{2}+P_{c}=\rho_{1}u_{1}^{2}=const\label{eq:mom}\end{equation}
Here $P_{c}$ is the CR pressure, $\rho_{1}$ and $u_{1}=V_{sh}$
are the far upstream density and flow speed. From these equation we
obtain the following equation for the shock front structure

\[
\left(\frac{\partial P_{c}}{\partial x}\right)^{2}+\frac{L_{B}}{\sigma P_{c}\left(1-P_{c}/P_{c2}\right)}\frac{\partial P_{c}}{\partial x}+\frac{\alpha}{\sigma}=0\]
where $P_{c2}=\left(6/7\right)\rho_{1}u_{1}^{2}$, $L_{B}=\bar{\kappa}_{B}/u_{1}$
and we assumed that thermal gas is not heated appreciably, so its
pressure can be neglected in eq.(\ref{eq:mom}) \citep{m97a}. The
shock solution can be obtained from the last equation in a closed
form as $x=x\left(P_{c}\right)$. Introducing a dimensionless coordinate
$z=xL_{B}/\sigma P_{c2}^{2}$ and $\Phi=2P_{c}/P_{c2}-1$ we can rewrite
the last equation describing the front transition as 

\[
\left(\frac{\partial\Phi}{\partial z}\right)^{2}+\frac{8}{1-\Phi^{2}}\frac{\partial\Phi}{\partial z}+4a=0\]
where the transition is governed by a single parameter $a=\alpha\sigma\left(P_{c2}/L_{B}\right)^{2}$.
Obviously, a smooth transition exists only for $0<a<4$. If $a\ll1$
$\Phi=-\tanh\left(az/2\right)$, so that the scale of the front is
the familiar $L_{B}=\bar{\kappa}_{B}/u_{1}$. If, however, $a\sim1$,
the scale of the shock (front) transition reduces to 

\[
L_{f}=\frac{\sigma}{4L_{B}}P_{c2}^{2}\simeq\frac{1}{4}\frac{L_{S}^{2}}{L_{B}}\left(\frac{P_{c2}}{\rho C_{s}^{2}}\right)^{2}.\]

\section{Conclusions}

We have obtained a one parameter family of smooth, strongly nonlinear
shock front transitions which are typically significantly shorter
than the conventional ($\sim\kappa/V_{sh}$) shock precursors. The
phenomenon in general is somewhat similar to the transport bifurcation
phenomenon, which is a subject of active research in magnetic fusion
(L-H bifurcations, \citep{Hinton91,DiamPRL94,MDtranspBif08}).

The critical parameter that governs such transitions (parameter $a$,
Sec.\ref{sec:Turbulent-CR-Front}) depends (through the growth rate
of the front supporting acoustic turbulence, $\gamma_{D}$) on the
thermal gas pressure inside the front, de facto on the turbulent heating
efficiency. The smooth transition exists for $a<4$, so that in the
case $a>4$, a gaseous subshock must form. This however, should not
result in a longer precursor. The determination of the heating efficiency,
and thus a more complete study of the shock fronts obtained in this
paper will be the subject of a separate publication.

Support by NASA under the Grants NNX 07AG83G and NNX 09AT94G as well
as by the Department of Energy, Grant No. DE-FG02-04ER54738 is gratefully
acknowledged.

\bibliographystyle{apj}
\bibliography{C:/TeX/BIBS/dsa,C:/TeX/BIBS/MALKOV,C:/TeX/BIBS/DSAobs,C:/TeX/BIBS/DSAshort,C:/TeX/BIBS/PlasmaDSA,C:/TeX/BIBS/tokamak}

\end{document}